\documentclass[letterpaper, 10 pt, conference]{ieeeconf}
\IEEEoverridecommandlockouts
\usepackage[T1]{fontenc}
\usepackage{graphicx}
\usepackage{amssymb}
\usepackage{amsmath}
\let\labelindent\relax
\usepackage{enumitem}
\usepackage{booktabs}
\usepackage{microtype}
\usepackage{subfigure}
\usepackage{url}

\newtheorem{remark}{Remark}

\begin{document}

\title{\LARGE \bf Trusted Polytopic Action Sets for Fast Planning in Underactuated Systems\looseness=-1}

\author{
Akshay Jaitly and 
Siavash Farzan
\thanks{Akshay Jaitly is with Onyx Robotics, 
Boston, MA 02210, USA (e-mail: akshay@onyx-robotics.com).}
\thanks{Siavash Farzan is with the Department of Electrical Engineering, California Polytechnic State University, San Luis Obispo, CA 93407, USA (e-mail: sfarzan@calpoly.edu).}
}

\maketitle
\thispagestyle{empty}
\pagestyle{empty}
 
\begin{abstract}
Underactuated systems pose a challenge for convex motion planning because their dynamically feasible motions lie on a manifold of trajectories in function space. Building on our earlier formulation of polytopic action sets (PAS), this paper presents a method for rapidly generating, online, trusted convex sets of short-horizon actions for underactuated and potentially nonlinear systems. Around a nominal trajectory, we construct local finite-dimensional action coordinates in which each parameter vector encodes a complete nearby motion through an affine trajectory map, rendering collision-avoidance and control bounds linear. To remain consistent with the nonlinear dynamics, we introduce a dynamics-violation metric and extract a trusted convex inner approximation using an IRIS-inspired inflation procedure directly in action space. The resulting PAS are reusable convex families of actions that can be queried and composed with linear programs, and a PAS-guided tree expansion treats nodes as composed reachable families rather than single trajectories, coupling local nonlinear fidelity with convex reuse for longer-horizon planning. The planner solves cluttered planar scenes in tens of milliseconds (14--78$\times$ faster than a kinodynamic RRT
baseline) and reduces terminal error on a nonlinear underactuated
benchmark by 26--86\% over sampling and NLP baselines.
\end{abstract}

% \begin{IEEEkeywords}
% Motion planning, convex optimization, constrained control, underactuated systems.
% \end{IEEEkeywords}

\section{Introduction}
\label{sec:introduction}

Convex representations have become a powerful tool in motion planning because they allow large classes of safety and trajectory constraints to be enforced with scalable optimization methods. In robotics, this is commonly realized by covering free or configuration space with convex regions and planning over those regions with Iterative Regional Inflation by Semidefinite programming (IRIS) and Graphs of Convex Sets (GCS) style methods \cite{Deits2015,Marcucci2023}. Representative examples include IRIS-style region inflation, fast sampling-based variants, and graph-based planning over families of convex sets \cite{Werner2026,Marcucci2024}. These methods have substantially expanded the range of problems that can be solved with linear, quadratic, and conic optimization.
Differentiable time-varying corridors offer a complementary route \cite{Jaitly2025,Arrizabalaga2024}, but none of these approaches address underactuated or nonlinear dynamics.\looseness=-1

\begin{figure}[t]
    \centering
    \includegraphics[trim={2.25cm 0 3cm 0},clip,width=0.7\columnwidth]{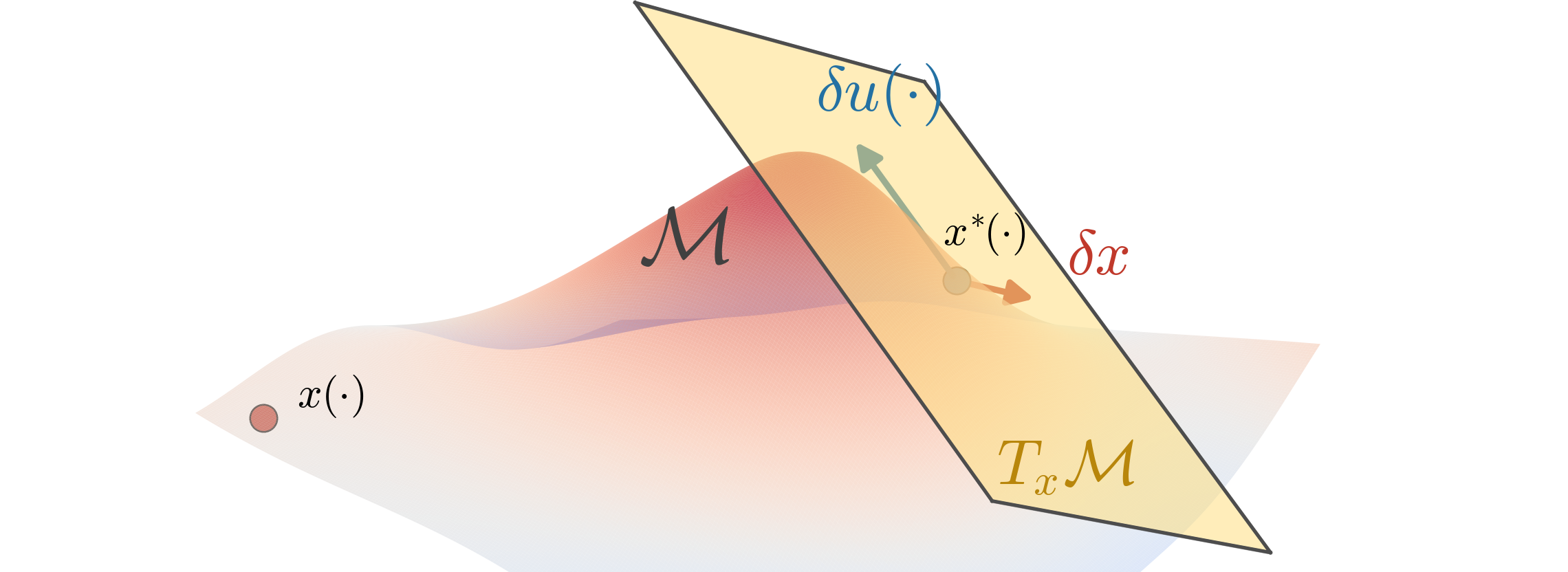} \\[2pt]
    \includegraphics[trim={2.25cm 0 1.25cm 0},clip,width=0.8\columnwidth]{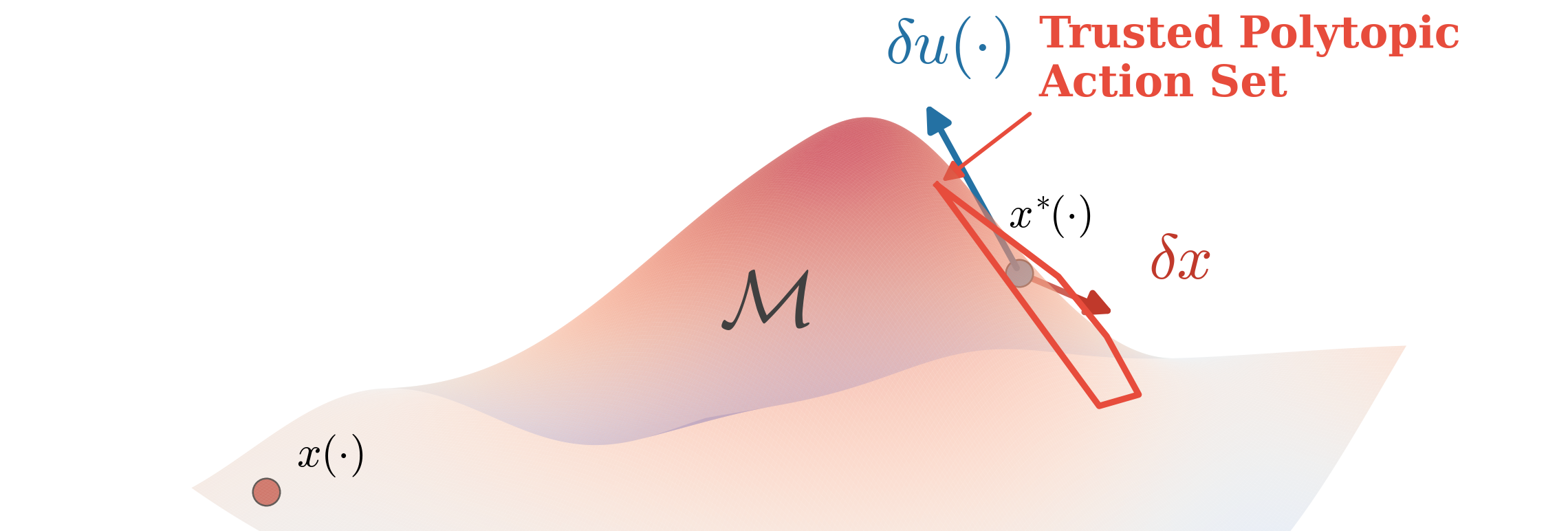}
    \vskip 3pt
    \caption{Geometric view of PAS construction. Top: a local LTV approximation around the nominal parameterizes nearby trajectories as affine functions of the action parameter $\gamma$. Bottom: in these action coordinates, a trusted PAS (red polytope) is extracted as a convex subset of the feasible-trajectory manifold. Here $\mathcal{M}$ is the dynamics-consistent trajectory manifold; $x^*(\cdot)$ the nominal trajectory and $x(\cdot)$ a neighbor on $\mathcal{M}$; $\delta x,\delta u(\cdot)$ the state/input deviations; and $T_{x^*}\mathcal{M}$ the tangent space, locally identified with the action-coordinate space $\gamma\in\mathbb{R}^{n_\gamma}$.}\looseness=-1
    \label{fig:manifold_geom}
\end{figure}

For \emph{underactuated} systems, the dominant source of nonconvexity is often the dynamics: feasible motions do not fill an open subset of trajectory space, but instead lie on a thin dynamics-consistent manifold.
Even a known collision-free corridor then contains only a small subset of dynamically realizable trajectories, which complicates any direct extension of convex free-space methods to kinodynamic planning.\looseness=-1

Sampling-based kinodynamic planners enforce dynamics feasibility by tree expansion with forward simulation or steering \cite{LaValle2001}. Related methods, including funnel libraries, reachable-set methods, and trajectory optimization, certify local tubes or compute locally feasible nominal motions for nonlinear systems \cite{Majumdar2017,Althoff2013,Murray1984,Tassa2014}. Complementary control-oriented approaches, including nonlinear and tube-based model-predictive control \cite{Mayne2000,Mayne2005} and control-barrier-function safe steering \cite{Ames2019}, also enforce dynamic feasibility online, but do so by solving an optimization problem at each step rather than by reusing precomputed motion families. Our objective is different. We seek reusable convex families of short-horizon trajectory parameters that can be queried and composed online with linear programs.\looseness=-1

In our earlier work~\cite{Jaitly2024}, we introduced \emph{polytopic action sets} (PAS) as convex subsets of a finite-dimensional motion-parameter space that encode entire families of short-horizon trajectories. Once available, PAS can be composed across time and queried via linear programs for long-horizon planning. For underactuated systems, however, feasible trajectories lie on a manifold in function space, and building convex subsets of that manifold directly is generally intractable.

This paper addresses that gap. The key idea is to move the convexification step from state space into a local \emph{action-coordinate space}: a finite-dimensional vector space whose points $\gamma$ parameterize complete short-horizon trajectories around a nominal motion through the affine map of Eq.~\eqref{eq:pas_affine}. Around a nominal feasible motion, we build a finite-dimensional affine parameterization of nearby trajectories via local linearization, within which collision and control constraints become linear inequalities. We then restrict this set to parameters for which the nonlinear dynamics are well approximated, using a trust region defined through a dynamics-violation metric. Since the trust region is nonconvex and available only through oracle queries, we compute a convex inner approximation by adapting IRIS-ZO~\cite{Werner2026} to the action-parameter space, which is amenable to batched GPU evaluation for online generation of trusted PAS.

Rather than convexifying the full nonlinear manifold globally, we construct local convex families of actions anchored to nominal motions and kept within a prescribed tolerance of the true dynamics. Each PAS is compact and reusable. It encodes an entire convex family of short-horizon trajectories, so the same PAS can be queried repeatedly (projected onto different targets, composed with other PAS, or post-processed by a trajectory-level quadratic program (QP)) without rebuilding any local model, using only linear programs.
We integrate these trusted PAS into a sampling-based tree expansion procedure. In contrast to classical kinodynamic Rapidly-exploring Random Tree (RRT), tree nodes represent \emph{composed convex families} of multi-segment trajectories obtained by enforcing continuity across a sequence of PAS, giving the planner access to reachable \emph{sets} rather than reachable points. Fig.~\ref{fig:manifold_geom} summarizes the geometric viewpoint underlying our construction.

The main contributions of this paper are:
\begin{itemize}
    \item[i.] A linearly constrained convex action parameterization that encodes sampled collision-avoidance and actuator-bound constraints as halfspace inequalities in a finite-dimensional trajectory-parameter space.
    \item[ii.] Online construction of \emph{trusted polytopic action sets} by intersecting the linear action set with a nonlinear dynamics-consistency trust region defined by a trajectory-defect metric.
    \item[iii.] Adaptation of IRIS-ZO to action-parameter space using batched oracle evaluations to compute a convex inner approximation of the trusted action set.
    \item[iv.] A PAS-guided tree expansion scheme that grows composed reachable families until the goal is reached.
\end{itemize}

\section{Polytopic Action Sets}
\label{sec:pas_background}

A \emph{Polytopic Action Set} (PAS) is a convex polytope $\mathcal{A}$ in a finite-dimensional parameter space $\mathbb{R}^{n_\gamma}$, where each point $\gamma \in \mathcal{A}$ encodes a short-horizon trajectory via an affine map:
\begin{equation}
    x(t;\gamma) = x^*(t) + R(t)\gamma, \quad u(t;\gamma) = u^*(t) + U(t)\gamma,
    \label{eq:pas_affine}
\end{equation}
for nominal trajectory $(x^*(t), u^*(t))$ and linear maps $R(t)$, $U(t)$ derived from the local dynamics (constructed in Section~\ref{sec:pas_generation}). The nominal action $\gamma = 0$ always encodes the nominal trajectory.\looseness=-1

\subsection{Reachable Sets}

\begin{figure}[t]
\centering
\subfigure[]{\includegraphics[trim={0 5pt 0 0},clip,width=0.53\columnwidth]{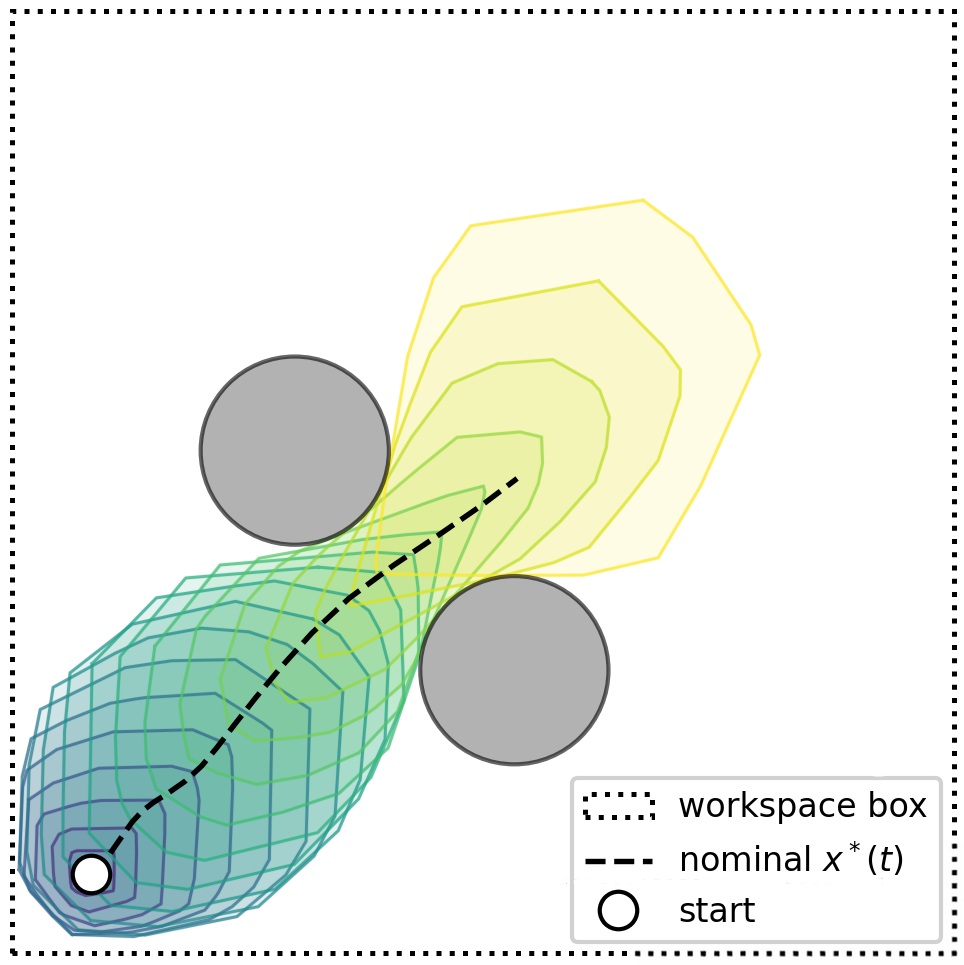}}
\hfill
\subfigure[]{\includegraphics[trim={35pt 20pt 32pt 20pt},clip,width=0.45\columnwidth]{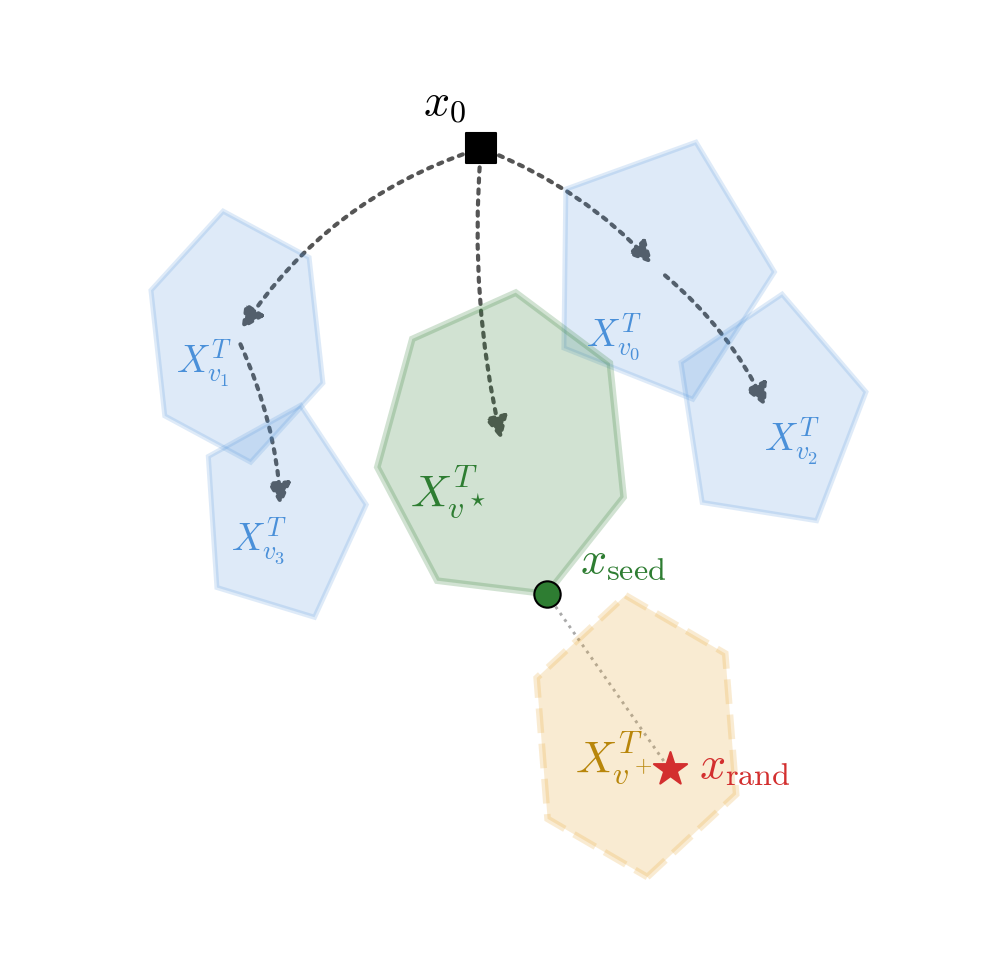}}
\caption{
(a) Reachable sets of a single PAS for the 2D double integrator, drawn at sampled time indices; the dashed curve is the nominal trajectory $x^*(t)$ about which the PAS is constructed, and sampled collision constraints enter as halfspaces separating the reachable sets from the obstacle.
(b) Composing PAS extends the reachable set toward the goal, the mechanism underlying our PAS-guided tree expansion (Sec.~\ref{sec:pas_tree_expansion}).}
\label{fig:pas_geometry}
\end{figure}

A key quantity for planning is the \emph{reachable set} at time $t$: the set of states attainable by some $\gamma \in \mathcal{A}$,
\begin{equation}
    \mathcal{X}(t) := \{ x^*(t) + R(t)\gamma \mid \gamma \in \mathcal{A} \}.
    \label{eq:reachable_set1}
\end{equation}
Because $\mathcal{A}$ is a polytope and $R(t)$ is linear, $\mathcal{X}(t)$ is itself a polytope. Fixing the initial state simply adds a linear constraint on $\gamma$. Querying membership, checking goal containment, or finding the closest reachable state to a target all reduce to linear programs (LPs) over $\gamma$. Fig.~\ref{fig:pas_geometry}a illustrates the forward reachable sets of a PAS for a double-integrator navigating obstacles, color-coded by time.

\subsection{Composability}

A central benefit of PAS is that they compose cleanly across segments while preserving convexity.
Given two sequential PAS $\mathcal{A}_j$ with trajectory maps
$x_{j}(t;\gamma_j)$, $j\,{\in}\,\{1,2\}$, the set of jointly feasible
parameter pairs satisfying inter-segment continuity,
\begin{equation}
    \mathcal{C} := \left\{ (\gamma_1, \gamma_2) \;\middle|\; \gamma_j \in \mathcal{A}_j,\; x_{1,N}(\gamma_1) = x_{2,0}(\gamma_2) \right\},
    \label{eq:composed_set_2seg}
\end{equation}
is a convex polyhedron, since all constraints are linear in $(\gamma_1, \gamma_2)$. The multi-segment composition and its convexity-preservation argument are developed in our earlier work~\cite{Jaitly2024}. This extends inductively as composing $\ell$ trusted PAS yields a single convex polyhedron whose feasible points are continuous multi-segment trajectories, over which post-hoc trajectory optimization (e.g., minimizing control effort) is a single convex program.\looseness=-1

\section{PAS Generation}
\label{sec:pas_generation}

We construct a trusted PAS around a nominal trajectory by building local action coordinates, imposing linear constraints, and refining with a nonlinear trust-region test (Fig.~\ref{fig:manifold_geom}).

\subsection{Local Trajectory Parameterization}
\label{subsec:local_parameterization}

Dynamically feasible motions of a nonlinear underactuated system lie on a dynamics-consistent manifold in function space. We construct \emph{local action coordinates} around a nominal feasible trajectory: a finite-dimensional parameterization of nearby motions in which constraints become linear.

Consider the nonlinear underactuated system
\begin{equation}
    \dot{x}(t) = f(x(t), u(t)),
    \;
    x(t) \in \mathbb{R}^{n_x},\;
    u(t) \in \mathbb{R}^{n_u},
    \;
    n_u {<} n_x.
    \label{eq:nonlinear_dynamics}
\end{equation}
Let $(x^*(t), u^*(t))$ be a nominal feasible trajectory over $t \in [0,T]$, where $T$ is the per-PAS planning horizon, discretized into $N$ steps of size $\Delta t = T/N$. Linearizing about this nominal gives a Linear Time-Varying (LTV) system with Jacobians
\[
    A(t) := \frac{\partial f}{\partial x}\bigg|_{(x^*(t),\,u^*(t))},
    \qquad
    B(t) := \frac{\partial f}{\partial u}\bigg|_{(x^*(t),\,u^*(t))},
\]
governing the evolution of deviations $\delta x(t) := x(t) - x^*(t)$ and $\delta u(t) := u(t) - u^*(t)$:
\begin{equation}
    \dot{\delta x}(t) = A(t)\,\delta x(t) + B(t)\,\delta u(t).
    \label{eq:ltv_dynamics}
\end{equation}
Integrating~\eqref{eq:ltv_dynamics} yields
\begin{equation}
    \delta x(t)
    = \Phi(t,0)\,\delta x_0
    + \int_0^t \Phi(t,s)\,B(s)\,\delta u(s)\,ds,
    \label{eq:ltv_solution}
\end{equation}
where the state transition matrix $\Phi(t,s)$ satisfies
$\frac{\partial}{\partial t}\Phi(t,s) = A(t)\,\Phi(t,s)$, $\Phi(s,s) = I$.
\looseness=-1 We parameterize the input deviation as $\delta u(t) = L(t)\,\eta$, where $L(t) \in \mathbb{R}^{n_u \times n_\eta}$ encodes a chosen function class (e.g., a piecewise-linear (PWL) profile with $M$ knots). Here $L(t)$ is a fixed user-supplied basis matrix and $\eta\in\mathbb{R}^{n_\eta}$ a static coefficient vector. For the PWL basis used here, $\eta$ collects the $M$ knot input values that $L(t)$ interpolates; \emph{knots} are within-horizon control degrees of freedom, distinct from the composed \emph{segments} (one PAS each) of Section~\ref{sec:pas_tree_expansion}.
Substituting gives
\begin{equation}
    \delta x(t) = \Phi(t,0)\,\delta x_0 + K(t)\,\eta,
    \label{eq:delta_x_affine}
\end{equation}
where $K(t) := \int_0^t \Phi(t,s)\,B(s)\,L(s)\,ds$ is the input-to-state map.
Collecting the free quantities into
\begin{equation}
    \gamma :=
    \begin{bmatrix} \delta x_0 \\ \eta \end{bmatrix}
    \in \mathbb{R}^{n_\gamma},
    \qquad
    n_\gamma = n_x + n_\eta,
    \label{eq:gamma_def}
\end{equation}
and defining
$R(t) := \begin{bmatrix} \Phi(t,0) & K(t) \end{bmatrix}$,
$U(t) := \begin{bmatrix} 0 & L(t) \end{bmatrix}$,
every perturbed trajectory is affine in $\gamma$ as in~\eqref{eq:pas_affine}. Setting $\gamma = 0$ recovers the nominal.

\begin{remark}
When $A,B$ are constant (the Linear Time-Invariant (LTI) case), the construction reduces to the standard controllability Gramian, so LTI is a special case. In practice $\Phi(t,0)$ and $K(t)$ are computed numerically by integrating the LTV dynamics on the discretization grid.
\end{remark}

A key advantage of this parameterization is that endpoint constraints are linear in $\gamma$. An initial condition $x(0) = \bar{x}_0$ requires $R(0)\,\gamma = \bar{x}_0 - x^*(0)$, and a terminal constraint $x(T) = \bar{x}_T$ becomes $R(T)\,\gamma = \bar{x}_T - x^*(T)$. Both are linear feasibility conditions on $\gamma$.

\subsection{Linear Collision and Control Constraints}
\label{subsec:linear_constraints}

The local coordinates let us enforce sampled geometric and actuator constraints directly in the parameter space $\gamma$.

Let $\mathcal{K}_{\mathrm{obs}} \subseteq \{0,\dots,N\}$ be the collision-checking indices and, for each $k$, let $P_k$ be a set of sampled forbidden points. For $p \in P_k$, the $\gamma$ that relate to trajectories that reach $p$ must be excluded. A separating half-space follows from any margin $0 < \epsilon_{p,k} < \|p-x_k^*\|_2^2$:
\begin{equation}
    \big(R_k^\top (p-x_k^*)\big)^\top \gamma
    \le
    \|p-x_k^*\|_2^2 - \epsilon_{p,k}.
    \label{eq:collision_halfspace}
\end{equation}
The nominal action $\gamma=0$ satisfies~\eqref{eq:collision_halfspace} by construction, while any $\gamma$ placing the state at $p$ violates it. Stacking over all $k$ and $p$ gives
\begin{equation}
    A_{\mathrm{col}} \gamma \le b_{\mathrm{col}}.
    \label{eq:stacked_collision_constraints}
\end{equation}
Let $u_{\min},u_{\max}\,{\in}\,\mathbb{R}^{n_u}$ denote componentwise actuator bounds. When the control parameterization $u(t)\,{=}\,L(t)\gamma_u$ defines a convex combination of known control points, we can enforce bounds on those control points to ensure feasibility of $u(t)$:
\begin{equation}
    A_u \gamma \le b_u.
    \label{eq:stacked_control_constraints}
\end{equation}

Combining~\eqref{eq:stacked_collision_constraints} and~\eqref{eq:stacked_control_constraints} yields
\begin{equation}
    \mathcal{A}_{\mathrm{lin}}
    :=
    \left\{
        \gamma \in \mathbb{R}^{n_\gamma}
        \;\middle|\;
        A_{\mathrm{col}} \gamma \le b_{\mathrm{col}},
        \;
        A_u \gamma \le b_u
    \right\}.
    \label{eq:A_lin}
\end{equation}
By construction $\gamma=0 \in \mathcal{A}_{\mathrm{lin}}$, so the nominal motion is retained, and $\mathcal{A}_{\mathrm{lin}}$ is a convex family of locally admissible actions under the linearized model. Fig.~\ref{fig:pas_geometry}a shows the reachable sets of such an action set for a double integrator: collision constraints project to hyperplanes separating the reachable tube from configurations in collision.

This suffices for LTI or LTV systems. Its limitation is that $\mathcal{A}_{\mathrm{lin}}$ does not guarantee consistency with the full nonlinear dynamics away from the nominal, which we address next.

\subsection{Dynamics-Consistency Trust Region}
\label{subsec:trust_region}

\begin{figure}
    \centering
    \includegraphics[width=0.625\columnwidth]{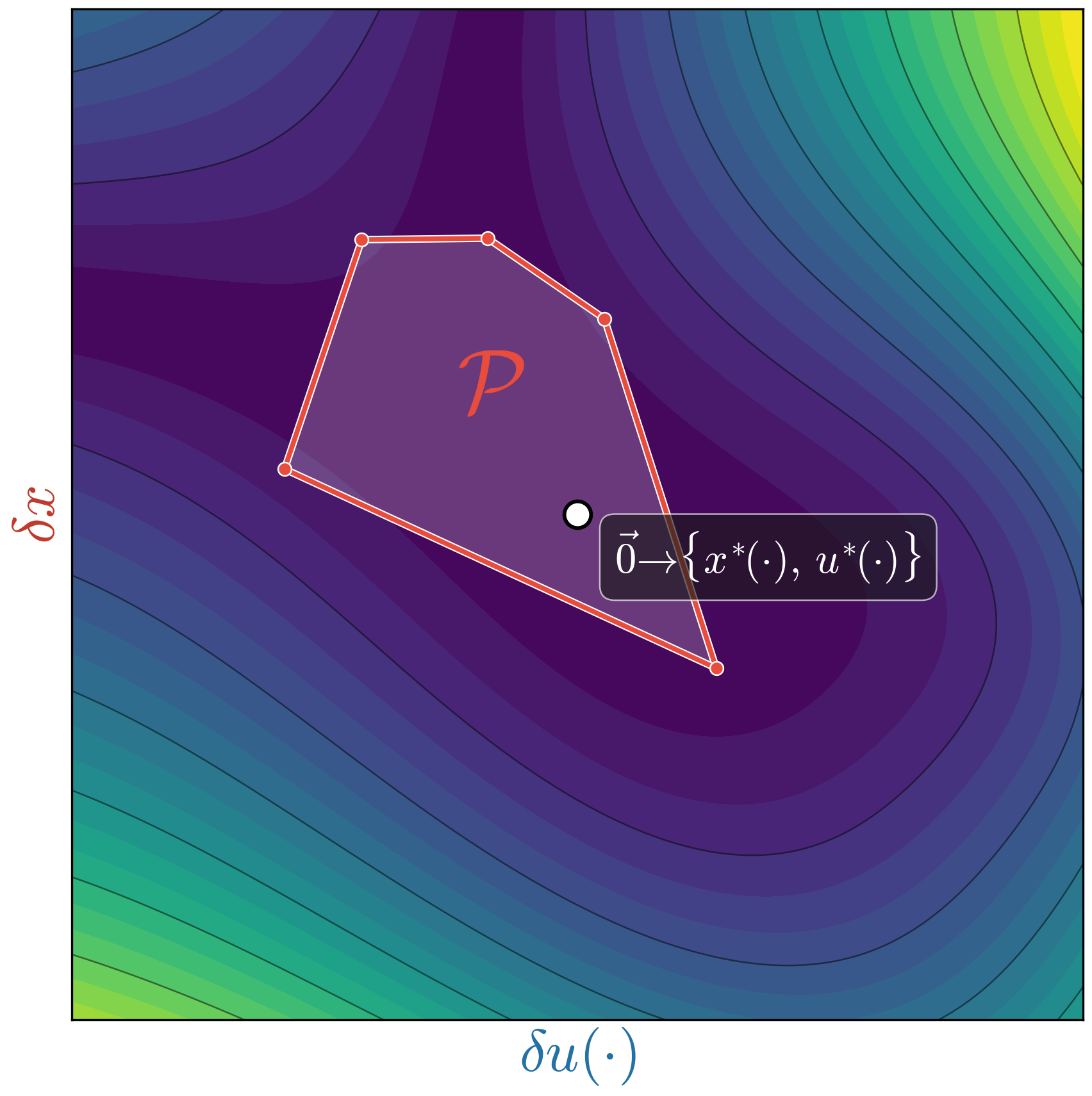}
    \caption{Dynamics-defect level sets $\zeta(\gamma)$ in action space. The trusted PAS (orange) lies inside the sublevel set $\zeta(\gamma) \leq \zeta_{\max}$.}
    \label{fig:tangent_distance}
\end{figure}

We measure departure from the nonlinear dynamics by a one-step defect on the discretization grid. At each index $k$ we compare the next state $x_{k+1}(\gamma)$ of the parameterized trajectory to a one-step integration of the true dynamics from $(x_k(\gamma),u_k(\gamma))$, and sum the squared residuals. This vanishes on the nominal and grows as the parameterization diverges from the true dynamics. Define the dynamics-violation functional
\begin{equation}
    \zeta(\gamma)
    {:=}
    \sum_{k=0}^{N-1}
    \left\|
        x_{k+1}(\gamma)
        {-}
        x_k(\gamma)
        {-}
        \Delta t\, f\!\big(x_k(\gamma),u_k(\gamma)\big)
    \right\|_2^2
    \label{eq:zeta_def}
\end{equation}
By construction $\zeta(0)=0$, since the nominal trajectory is dynamically feasible. As $\|\gamma\|$ grows, the linear approximation becomes less accurate and $\zeta(\gamma)$ increases, as illustrated in Fig.~\ref{fig:tangent_distance}.

The \emph{trust region} in parameter space is the sublevel set
\begin{equation}
    \mathcal{R}_{\mathrm{trust}}(\zeta_{\max})
    :=
    \left\{
        \gamma \in \mathbb{R}^{n_\gamma}
        \;\middle|\;
        \zeta(\gamma) \le \zeta_{\max}
    \right\},
    \label{eq:trust_region_def}
\end{equation}
where $\zeta_{\max} > 0$ is a user-chosen tolerance. Fig.~\ref{fig:tangent_distance} illustrates this sublevel set together with the trusted PAS it contains.
Smaller values yield more conservative action sets that stay closer to the nominal dynamics manifold, larger values admit broader exploration at the cost of reduced local model fidelity. The set $\mathcal{R}_{\mathrm{trust}}(\zeta_{\max})$ is generally nonconvex. Our target is therefore the nonconvex intersection
\begin{equation}
    \Omega_{\mathrm{trust}}
    :=
    \mathcal{A}_{\mathrm{lin}}
    \cap
    \mathcal{R}_{\mathrm{trust}}(\zeta_{\max}).
    \label{eq:omega_trust}
\end{equation}

\subsection{Zeroth-Order Action-Space Inner Approximation}
\label{subsec:zo_inflation}

We compute a convex inner approximation of $\Omega_{\mathrm{trust}}$ by adapting IRIS-ZO~\cite{Werner2026} to the action-parameter space $\gamma$ rather than configuration space.
Define the feasibility oracle
\begin{equation}
    g(\gamma)
    =
    \begin{cases}
        1, & \text{if } \zeta(\gamma)\le \zeta_{\max}, \\
        0, & \text{otherwise,}
    \end{cases}
    \label{eq:feasibility_oracle}
\end{equation}
so that $g(\gamma)=1$ if and only if the trajectory encoded by $\gamma$ lies within the prescribed dynamics tolerance. Evaluating $g(\gamma)$ requires only forward evaluation of the parameterized trajectory and residual, no gradients or nonlinear solves are needed, which makes the construction compatible with large batched evaluations.

We compute a convex inner approximation of $\Omega_{\mathrm{trust}}$ by adapting the cutting stage of IRIS-ZO to the action-parameter space $\gamma$. 
Starting from $\mathcal{A}^{(0)} := \mathcal{A}_{\mathrm{lin}}$, which contains $\gamma=0$ by construction, we iteratively refine the polytope, with the goal of separating the interior of the polytope from the space where the linearized model is inaccurate. At each iteration, we sample a batch of candidate parameters from the current polytope and evaluate $g(\gamma)$ in parallel. For each infeasible sample, we bisect along the ray from $0$ to $\gamma$ to localize the boundary of  $\mathcal{R}_{\mathrm{trust}}$.
After $M_{\text{iter}}$ rounds, the resulting polytope is the final \emph{trusted polytopic action set}:
\begin{equation}
    \mathcal{A}_{\mathrm{trust}}
    :=
    \mathcal{A}^{(M_{\mathrm{iter}})},
    \qquad
    \mathcal{A}_{\mathrm{trust}} \subseteq \mathcal{A}_{\mathrm{lin}}.
    \label{eq:A_trust_def}
\end{equation}
By construction, $\mathcal{A}_{\mathrm{trust}}$ is anchored at the dynamically feasible nominal $\gamma=0$, and each accumulated halfspace removes a sampled infeasible direction, so the action set is concentrated around the nominal trajectory. Because the refinement is zeroth-order and sample-based, $\mathcal{A}_{\mathrm{trust}}$ is an \emph{approximate} convex inner approximation of $\Omega_{\mathrm{trust}}$ rather than a certified one. The nonconvex boundary of $\mathcal{R}_{\mathrm{trust}}$ is resolved up to the sampling resolution, and the approximation tightens as the number of cutting rounds $M_{\mathrm{iter}}$ and action-space samples per round increase. The residual approximation gap, and the conservatism with which dynamically feasible motions are excluded, depend on the local geometry of $\mathcal{R}_{\mathrm{trust}}$ around the nominal. They are more pronounced for strongly nonlinear systems or poorly conditioned LTV models, and negligible in the LTI case, where the model is exact and no refinement is needed. We revisit this
in Section~\ref{sec:results}.\looseness=-1

\section{PAS-Guided Tree Expansion}
\label{sec:pas_tree_expansion}

We use trusted PAS as primitives in a sampling-based tree expansion. Each tree node represents a \emph{composed convex family} of multi-segment motions, giving the planner access to reachable \emph{sets} rather than point states, as shown in Fig.~\ref{fig:pas_geometry}b.

\subsection{Node Representation by Composed Action Sets}
\label{subsec:node_representation}

Let $\mathcal{T} = (V,E)$ be the search tree rooted at $x_{\mathrm{init}}$, each edge one trusted PAS. A node $v$ represents a sequence $\{\mathcal{A}_1,\dots,\mathcal{A}_{\ell(v)}\}$ of action sets, each representing a segment $j$, 
\begin{equation}
    \mathcal{A}_j = \{\gamma_j \in \mathbb{R}^{n_\gamma} \mid A_j \gamma_j \le b_j\},
    \label{eq:segment_pas}
\end{equation}
with local affine trajectory maps $x_{j,k}(\gamma_j) = x_{j,k}^* + R_{j,k}\gamma_j$, $u_{j,k}(\gamma_j) = u_{j,k}^* + U_{j,k}\gamma_j$.

Stacking segment parameters into $\Gamma_v := [\gamma_1^\top \;\cdots\; \gamma_{\ell(v)}^\top]^\top$, the composed feasible set is
\begin{align}
    \mathcal{C}_v
    {:=}
    \Big\{
        \Gamma_v \;\Big|\;
        &A_j\gamma_j \le b_j,\quad j=1,\dots,\ell(v), \label{eq:composed_set}\\[-2pt]
        &x_{1,0}(\gamma_1)=x_{\mathrm{init}}, \nonumber\\[-2pt]
        &x_{j,N}(\gamma_j)=x_{j+1,0}(\gamma_{j+1}),\, j=1,\dots,
        \,\ell(v)-1 \nonumber
    \Big\}.
\end{align}
All constraints are linear in $\Gamma_v$, so $\mathcal{C}_v$ is a convex polyhedron whenever nonempty. The terminal reachable set of node $v$ is
\begin{equation}
    X_v^T
    :=
    \left\{
        x_{\ell(v),N}(\gamma_{\ell(v)})
        \;\middle|\;
        \Gamma_v \in \mathcal{C}_v
    \right\},
    \label{eq:terminal_reachable_set}
\end{equation}
a convex set of terminal states reachable by a continuous sequence of trusted local actions. The superscript in $X_v^T$ denotes \emph{terminal}, not the horizon length $T$.

\subsection{Expansion Procedure}
\label{subsec:pas_expansion}

The planner grows the tree until some node's reachable set contains $x_{\mathrm{goal}}$. Each expansion proceeds as follows.

\paragraph*{1) Sample and select a node}
Draw a random target $x_{\mathrm{rand}}$ (optionally biased toward $x_{\mathrm{goal}}$). Select a node $v^\star$
\begin{equation}
    v^\star = \arg\min_{v \in V} \left\| W(\hat{x}_v - x_{\mathrm{rand}}) \right\|_2,
    \label{eq:node_prescore}
\end{equation}
where $W$ is a diagonal weight matrix masking irrelevant state components, and $\hat{x}_v$ is a characteristic state representing node $v$. This reduces node selection to a nearest-neighbor lookup.

\paragraph*{2) Project into the reachable set}
Solve one LP to find the closest point (in a weighted $\ell_\infty$-norm) in $X_{v^\star}^T$ to $x_{\mathrm{rand}}$:
\begin{equation}
    \min_{\Gamma_{v^\star},\,r}\; r
    \;
    \text{s.t.}\;
    \Gamma_{v^\star} \in \mathcal{C}_{v^\star},\;
    \bigl\| W(x_{v^\star}^T(\Gamma_{v^\star}) \,{-}\, x_{\mathrm{rand}}) \bigr\|_\infty \,{\le}\, r.
\label{eq:projection_lp}
\end{equation}
The optimizer yields an $x_{\mathrm{seed}}$ we can reach from $x_{\mathrm{init}}$.

\paragraph*{3) Steer and build a new PAS}
From $x_{\mathrm{seed}}$, a short-horizon steering routine produces a nominal trajectory toward $x_{\mathrm{rand}}$. Applying Section~\ref{sec:pas_generation} yields a new trusted PAS $\mathcal{A}_{\mathrm{new}}$. A child node is formed by appending $\mathcal{A}_{\mathrm{new}}$ and enforcing continuity at $x_{\mathrm{seed}}$:
\begin{equation}
\begin{aligned}
    \mathcal{C}_{v^+}
    :=
    \Big\{
        (\Gamma_{v^\star},\gamma_{\mathrm{new}})
        \;\Big|\;
        &\Gamma_{v^\star}\in\mathcal{C}_{v^\star},\;
        A_{\mathrm{new}}\gamma_{\mathrm{new}} \le b_{\mathrm{new}},\\[-2pt]
        &x_{v^\star}^T(\Gamma_{v^\star})
        = x_{\mathrm{new},0}(\gamma_{\mathrm{new}})
    \Big\}.
\end{aligned}
\label{eq:child_set}
\end{equation}
This set is nonempty by construction since $\Gamma_{v^\star}^\star$ and $\gamma_{\mathrm{new}}\,{=}\,0$ both reach $x_{\mathrm{seed}}$.
Fig.~\ref{fig:pas_geometry}b shows the resulting reachable set, newly formed to contain $x_{\mathrm{rand}}$.

\paragraph*{4) Goal test and caching}
Test whether $x_{\mathrm{goal}} \in X_{v^+}^T$ by solving a linear feasibility program. If feasible, recover the complete multi-segment plan and terminate. If not, the closest reachable point to $x_{\mathrm{goal}}$ within $X_{v^+}^T$ becomes the cached $\hat{x}_{v^+}$ for future node selection. Return to step~1.

\section{Results and Discussion}
\label{sec:results}

\subsection{Simulation Setup}
\label{subsec:experimental_setup}

We evaluate PAS-RRT on two systems: (i)~a 2D double integrator ($n_x\,{=}\,4$, $n_u\,{=}\,2$, LTI) navigating obstacle fields, and (ii)~a cartpole swing-up ($n_x\,{=}\,4$, $n_u\,{=}\,1$), for which PAS are constructed from local LTV approximations of the underactuated nonlinear dynamics. The double integrator uses a $10{\times}10$\,m workspace with circular obstacles of radius $r\,{=}\,0.6$\,m, $\Delta t\,{=}\,0.1$\,s, and $u\,{\in}\,[-5,5]^2$\,N. The cartpole uses $\Delta t\,{=}\,0.015$\,s,
$u\,{\in}\,[-10,10]$\,N, a cart-position range $x\,{\in}\,[-3,3]$\,m, and
$W\,{=}\,\mathrm{diag}(0,1,1,1)$, so the swing-up goal does not constrain the
final cart position.
Control inputs are parameterized with a piecewise-linear (PWL) basis with $M$ knots. For cartpole, the trust-region test uses $\zeta_{\max}\,{=}\,0.05$. For the LTI system, no trust-region refinement is needed because the local model is exact.

The nominal trajectory seeding each PAS is produced by a sampling-based steering routine: random PWL control sequences over the $M$ knot times are rolled out under the true dynamics and ranked by $W$-masked terminal distance to the target. The LTI system draws $512$ Gaussian knot samples ($\sigma{=}3$\,N, clipped to the actuator bounds) and the cartpole $1024$ ($\sigma{=}5$\,N). The IRIS-ZO refinement uses $256$ action-space samples and $12$ bisection steps per round, with $M_{\mathrm{iter}}\,{=}\,1$ (cartpole) and $0$ (LTI). Sampling and rollouts run batched on an NVIDIA RTX~5060 GPU, planning LPs use HiGHS and the smoothing QP uses OSQP on the CPU. The line-search Nonlinear Programming (NLP) baseline runs a finite-difference Gauss--Newton search with $8$ restarts, treating $\zeta\,{\le}\,\zeta_{\max}$ as a soft penalty.

\subsection{LTI Obstacle Avoidance}
\label{subsec:lti_results}

\begin{figure}[t]
\centering
\includegraphics[trim={15pt 10pt 15pt 1cm}, clip,width=\columnwidth]{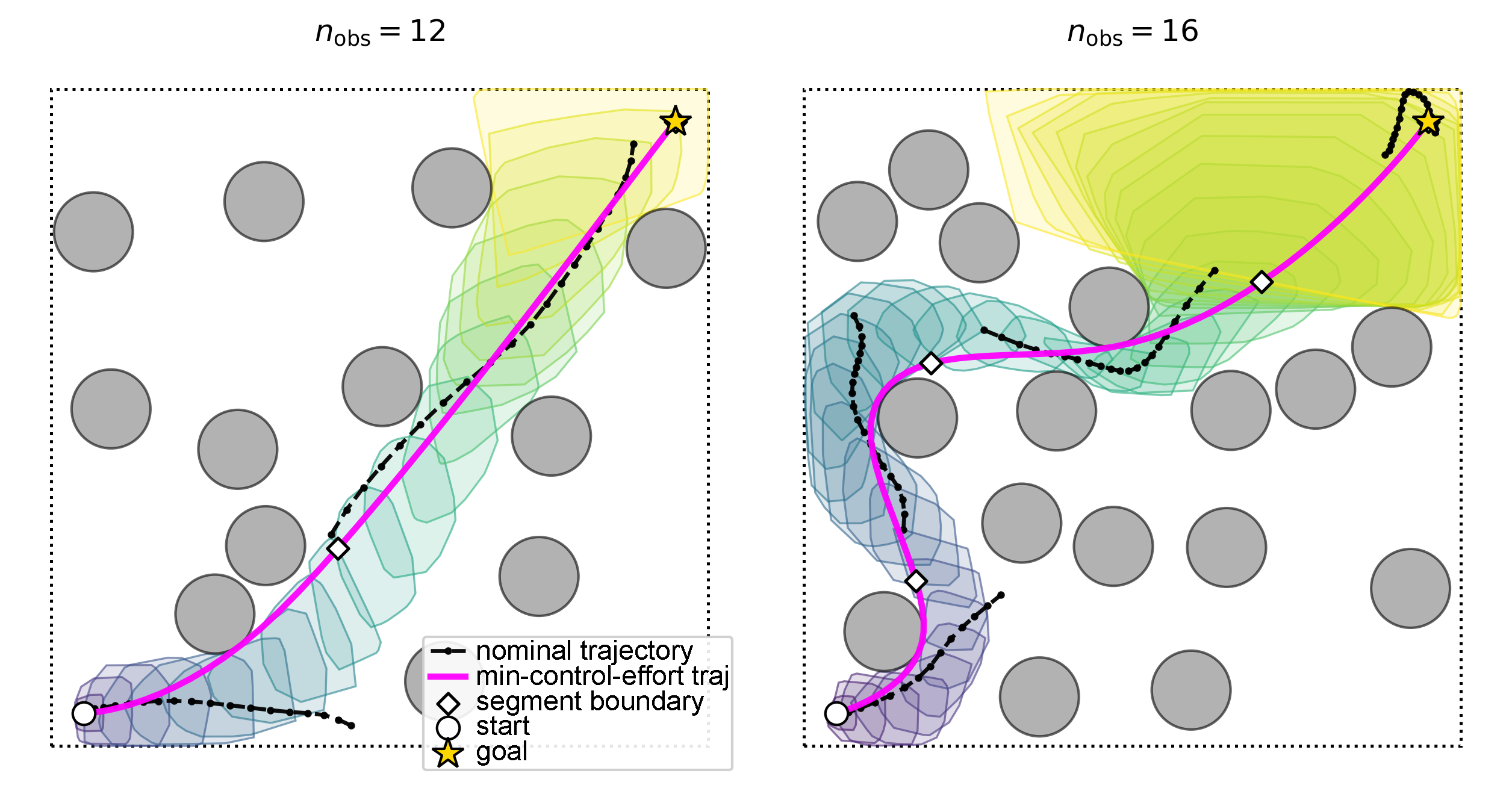}
\caption{Representative PAS-RRT solutions at 12 obstacles (left) and 16 obstacles (right). Magenta: final trajectory; colored tubes: forward reachable sets; dashed: nominal steering rollouts.}
\label{fig:obstacle_gallery}
\vspace{-5pt}
\end{figure}

\begin{figure}[t]
  \centering
  \includegraphics[width=\linewidth]{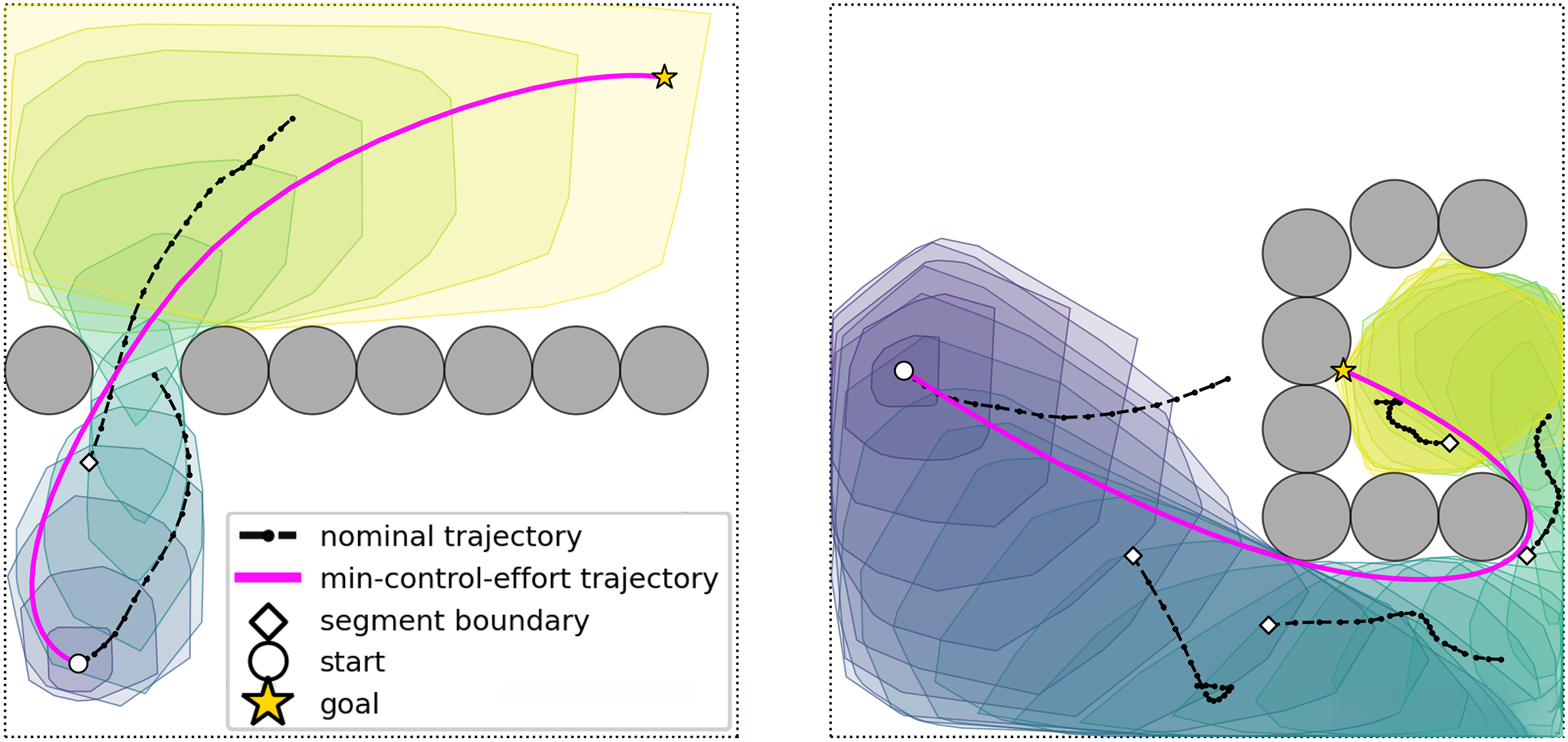}
  \caption{Scenes with non-trivial topology: a narrow passage (left) and a bug-trap enclosing the goal (right). PAS-RRT solves both; trajectories are minimum-effort QP solutions.}
  \label{fig:deadend}
\end{figure}

We sweep PWL knots $M\,{\in}\,\{5,10,15,20\}$, horizon steps $N\,{\in}\,\{10,20,30\}$, and obstacle count $n_{\mathrm{obs}}\,{\in}\,\{4,\dots,25\}$ (occupancy ${\approx}\,4.5$--$28\%$), with 50 randomized trials per configuration (3,600 total). PAS-RRT succeeds in all trials (500-expansion budget, goal tolerance $0.5$\,m). 
Representative solutions at intermediate densities appear in Fig.~\ref{fig:obstacle_gallery}, with reachable tubes widening at intermediate times and contracting near constraints.
The upper block of Table~\ref{tab:lti-occupancy} reports planning cost versus occupancy at fixed discretization ($M\,{=}\,10$, $N\,{=}\,20$): runtime grows smoothly from $9.6$\,ms at ${\approx}\,4.5\%$ to $127$\,ms at ${\approx}\,28\%$, with $2.5$--$4.9$ composed segments (the diamonds in Figs.~\ref{fig:obstacle_gallery}--\ref{fig:deadend} mark \emph{segment boundaries}, where consecutive PAS are joined by the continuity constraints of~\eqref{eq:composed_set}).\looseness=-1

Extending the sweep into the high-occupancy regime (lower block of Table~\ref{tab:lti-occupancy}), success holds at $100\%$ up to ${\approx}\,32\%$ occupancy, then degrades to $92\%$ near $41\%$ and $82\%$ at $63\%$. These failures are workspace-connectivity limited: the free passages approach the planner's collision margin, so some instances admit no margin-respecting corridor, resembling a denser variant of Fig.~\ref{fig:obstacle_gallery}.

Against a kinodynamic RRT baseline (OMPL) on the same scenes, under matched control bounds and collision margins, PAS-RRT is $14$--$78\times$ faster (e.g., ${\approx}\,67$\,ms versus ${\approx}\,2.0$\,s at $20$ obstacles) at comparable final path cost, by replacing per-node forward simulation with convex reuse of composed action sets.\looseness=-1

To exercise local-minimum escape, we add two scenes with non-trivial topology (Fig.~\ref{fig:deadend}): a narrow passage with a single gap, and a bug-trap in which a U-shaped obstacle wall encloses the goal. PAS-RRT reaches the goal in all $50$ trials of each. The bug-trap takes roughly an order of magnitude more expansions than the narrow-passage scene ($105$ vs. $10$ expansions) as the tree backs out of the cul-de-sac. A minimum-control-effort QP solved over the composed set yields the trajectories in Fig.~\ref{fig:deadend}.

\begin{table}[b]
\centering
\footnotesize
\caption{LTI obstacle avoidance: planning cost vs.\ workspace \\[-1pt] occupancy ($M\,{=}\,10$, $N\,{=}\,20$, 50 trials per row).\\[-1pt] Lower block: high-occupancy regime.}
\label{tab:lti-occupancy}
\setlength{\tabcolsep}{4.5pt}
\renewcommand{\arraystretch}{0.9}
\begin{tabular}{rrrrrr}
\toprule
$n_{\mathrm{obs}}$ & Occ.\,[\%] & Success\,[\%] & Expansion Iters & Plan [ms] & Seg. \\
\midrule
4  & 4.5  & 100 & 5.8  & 9.6   & 2.5 \\
8  & 9.0  & 100 & 10.5 & 18.3  & 3.0 \\
12 & 13.6 & 100 & 13.2 & 32.5  & 3.3 \\
16 & 18.1 & 100 & 20.0 & 40.8  & 3.8 \\
20 & 22.6 & 100 & 38.6 & 66.5  & 4.4 \\
25 & 28.3 & 100 & 59.8 & 127.0 & 4.9 \\
\midrule
36 & 40.7 & 92 & 81 & 305 & 6.0 \\
48 & 54.3 & 90 & 112 & 547 & 5.0 \\
56 & 63.3 & 82 & 105 & 552 & 5.0 \\
\bottomrule
\end{tabular}
\end{table}

\subsection{Cartpole Swing-Up with Local LTV PAS} \label{subsec:ltv_results}

The cartpole swing-up ($\theta\,{=}\,\pi {\to} 0$) tests PAS-RRT on nonlinear underactuated dynamics. We vary the PWL parameterization ($M$ knots), horizon length ($T$), and the number of cutting rounds $M_{\mathrm{iter}}$, and report medians over 20 trials per configuration.

\begin{figure}[t]
\centering
\includegraphics[width=\columnwidth]{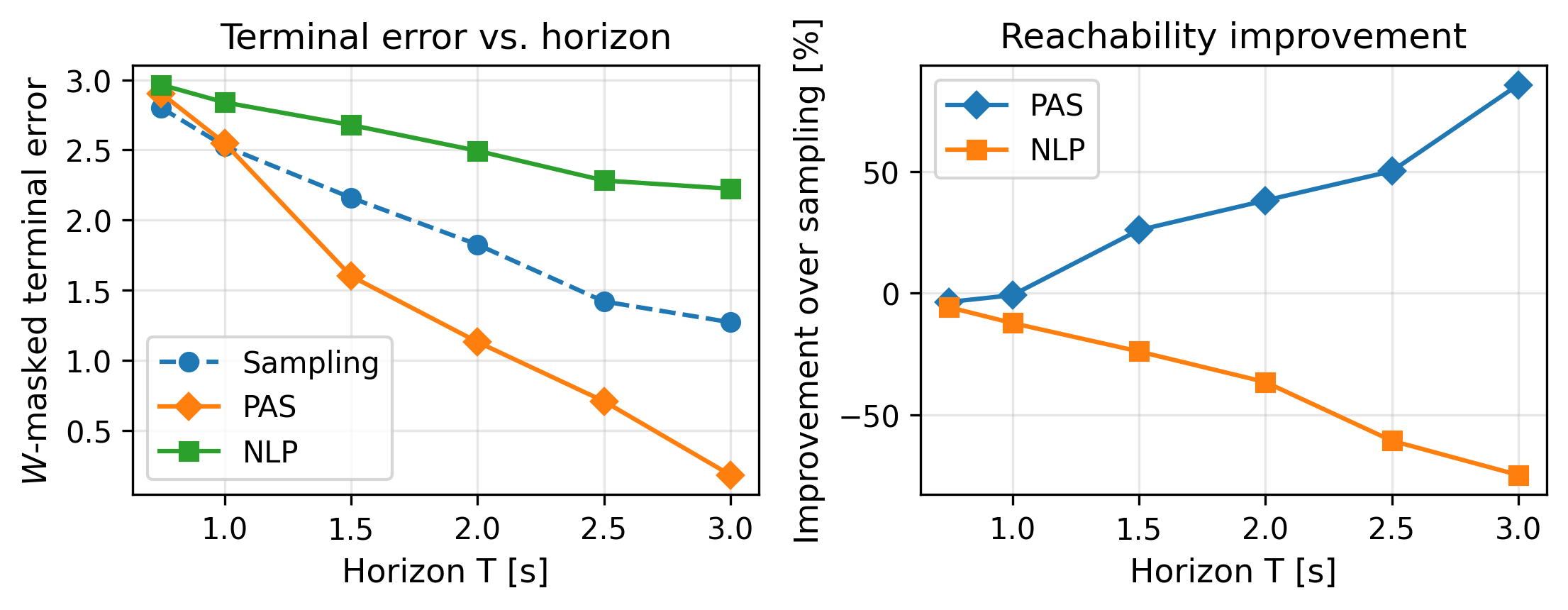}
\caption{Single-PAS cartpole goal reachability ($M\,{=}\,40$, $\zeta_{\max}\,{=}\,0.05$). Left: $W$-masked terminal error vs.\ horizon; right: relative improvement over direct sampling. PAS outperforms sampling and the NLP baseline.}
\label{fig:pas_vs_nlp}
\vspace{-5pt}
\end{figure}

A single steering rollout reaches the goal in no tested configuration: even at $T\,{=}\,3.0$\,s, $M\,{=}\,20$, the best of $1{,}024$ samples leaves a $W$-masked terminal error well above tolerance. PAS instead recovers a goal-reaching action through the goal-projection LP over a convex family of trajectories. As shown in Fig.~\ref{fig:pas_vs_nlp}, for $M\,{=}\,40$ the PAS solution reduces this error by $26$--$86\%$ over direct sampling for $T\,{\in}\,[1.5,3.0]$\,s, whereas the line-search NLP baseline does not improve on sampling: it applies $\zeta\,{\le}\,\zeta_{\max}$ as a soft penalty, so its iterates can drift outside the trusted region, and it returns a single point rather than a queryable family.

Table~\ref{tab:cartpole-sweep} summarizes planning performance. Shorter horizons require more expansions and larger tree depth, while longer horizons ($T\,{\geq}\,3.0$\,s) solve in $1$--$2$ expansions at depth $1$--$2$, with runtime dominated by PAS construction. Raising the number of cutting rounds $M_{\mathrm{iter}}$ mainly adds halfspaces and LP cost. The per-step dynamics defect stays small at longer horizons (${\leq}\,0.006$ for $T\,{\geq}\,3.0$\,s), the largest values arising in the strongly nonlinear swing-through regime.

Sweeping $\zeta_{\max}$ from $0.005$ to $1.0$ on a representative configuration, planner success stays at $100\%$ and planning time near $30$\,ms, while the per-step residual of the returned plan grows monotonically from ${\approx}\,2{\times}10^{-3}$ to ${\approx}\,1{\times}10^{-2}$. The tolerance trades fidelity against exploration smoothly, without a sharp threshold.\looseness=-1

Under control bounds and a cart-position range matched to PAS-RRT, the kinodynamic RRT baseline finds no swing-up within a $30$\,s budget. With the cart-position limits relaxed it succeeds, but ${\approx}\,15$--$20\times$ slower than PAS-RRT, which swings up in ${\approx}\,40$--$70$\,ms under the original constraints.
The composed PAS provides a corridor of dynamically consistent trajectories connecting start to goal, with the backward reachable set contracting toward the upright equilibrium.\looseness=-1

\begin{table}[t]
\centering
\footnotesize
\caption{PAS-RRT cartpole swing-up results ($\Delta t{=}0.015$\,s, 20 trials, \\[-2pt] median). Resid.: avg.\ per-step dynamics defect along solution.}
\label{tab:cartpole-sweep}
\setlength{\tabcolsep}{3.5pt}
\renewcommand{\arraystretch}{0.9}
\begin{tabular}{rrr|rr|rrr|r}
\toprule
$M$ & $T$ & $M_{\mathrm{iter}}$ & Expansion Iters & Depth & Total & Build & LP & Resid. \\
    & [s] &     &       &       & [ms]  & [ms]  & [ms] &  \\
\midrule
\multicolumn{9}{c}{\emph{Shorter horizon}} \\
\midrule
20 & 0.75 & 1 & 41.5 & 5.0 & 1297 & 8.1  & 21.2 & 0.016 \\
30 & 1.50 & 1 & 5.5  & 2.0 & 184  & 16.3 & 18.8 & 0.024 \\
\midrule
\multicolumn{9}{c}{\emph{Longer horizon}} \\
\midrule
10 & 3.00 & 1 & 1.0 & 1.0 & 26  & 19.0 & 3.7  & 0.001 \\
20 & 3.00 & 1 & 1.5 & 1.0 & 44  & 24.9 & 6.9  & 0.006 \\
20 & 3.00 & 5 & 2.0 & 1.5 & 154 & 88.6 & 10.5 & 0.003 \\
20 & 4.00 & 1 & 1.0 & 1.0 & 42  & 28.5 & 6.5  & 0.006 \\
\bottomrule
\end{tabular}
\end{table}

An open-source implementation of the proposed method is available at: \\
{\small\url{github.com/akshay5312/paamp_underactuated}}

\section{Conclusions}
\label{sec:conclusion}

This paper presented a method for rapidly generating trusted polytopic action sets for control-constrained systems. The key findings are: (i) a single action-space cutting round ($M_{\mathrm{iter}}\,{=}\,1$) is usually sufficient, as additional rounds add LP overhead without reducing tree expansions; (ii) the per-PAS horizon $T$ is the dominant planning parameter, with longer horizons reducing tree depth to as few as one or two expansions; and (iii) the convex-family structure of PAS recovers goal-reaching actions through the goal-projection LP \eqref{eq:projection_lp} that no single steering rollout, and the NLP baseline, attains.\looseness=-1

The trusted action set is an \emph{approximate} convex inner approximation: as the IRIS-ZO refinement is zeroth-order and sample-based, its quality improves with the cutting budget, and the residual gap and the conservatism with which feasible motions are excluded depend on system nonlinearity and on the conditioning of the local model at the nominal. The tolerance $\zeta_{\max}$ is well-conditioned over the swept range. The method remains inherently local: performance degrades when nominal rollouts are poor and, as the high-occupancy and dead-end studies show, in workspaces whose feasible passages are narrower than the planner's collision margin. Sampled collision constraints are likewise conservative with respect to the discretization. Future work will focus on adaptive trust-region selection, less conservative continuous-time safety certificates, and tighter coupling between PAS generation and steering.\looseness=-1
 
\bibliographystyle{IEEEtran}
\bibliography{bibtex}

\begin{thebibliography}{10}
\providecommand{\url}[1]{#1}
\csname url@samestyle\endcsname
\providecommand{\newblock}{\relax}
\providecommand{\bibinfo}[2]{#2}
\providecommand{\BIBentrySTDinterwordspacing}{\spaceskip=0pt\relax}
\providecommand{\BIBentryALTinterwordstretchfactor}{4}
\providecommand{\BIBentryALTinterwordspacing}{\spaceskip=\fontdimen2\font plus
\BIBentryALTinterwordstretchfactor\fontdimen3\font minus \fontdimen4\font\relax}
\providecommand{\BIBforeignlanguage}[2]{{%
\expandafter\ifx\csname l@#1\endcsname\relax
\typeout{** WARNING: IEEEtran.bst: No hyphenation pattern has been}%
\typeout{** loaded for the language `#1'. Using the pattern for}%
\typeout{** the default language instead.}%
\else
\language=\csname l@#1\endcsname
\fi
#2}}
\providecommand{\BIBdecl}{\relax}
\BIBdecl

\bibitem{Deits2015}
R.~Deits and R.~Tedrake, \emph{Computing Large Convex Regions of Obstacle-Free Space Through Semidefinite Programming}.\hskip 1em plus 0.5em minus 0.4em\relax Cham: Springer International Publishing, 2015, pp. 109--124.

\bibitem{Marcucci2023}
T.~Marcucci, M.~Petersen, D.~von Wrangel, and R.~Tedrake, ``Motion planning around obstacles with convex optimization,'' \emph{Science Robotics}, vol.~8, no.~84, p. 7843, 2023.

\bibitem{Werner2026}
P.~Werner, T.~Cohn, R.~H. Jiang, T.~Seyde, M.~Simchowitz, R.~Tedrake, and D.~Rus, ``Faster algorithms for growing collision-free convex polytopes in robot configuration space,'' 2024, arXiv:2410.12649.

\bibitem{Marcucci2024}
T.~Marcucci, J.~Umenberger, P.~Parrilo, and R.~Tedrake, ``Shortest paths in graphs of convex sets,'' \emph{SIAM Journal on Optimization}, vol.~34, no.~1, pp. 507--532, 2024.

\bibitem{Jaitly2025}
A.~Jaitly, J.~Arrizabalaga, and G.~Li, ``Trajectory planning using safe ellipsoidal corridors as projections of orthogonal trust regions,'' 2025, arXiv:2509.19734.

\bibitem{Arrizabalaga2024}
J.~Arrizabalaga, Z.~Manchester, and M.~Ryll, ``Differentiable collision-free parametric corridors,'' in \emph{2024 IEEE/RSJ International Conference on Intelligent Robots and Systems (IROS)}, 2024, pp. 1839--1846.

\bibitem{LaValle2001}
S.~M. LaValle and J.~J. KuffnerJr., ``Randomized kinodynamic planning,'' \emph{The International Journal of Robotics Research}, vol.~20, no.~5, pp. 378--400, 2001.

\bibitem{Majumdar2017}
A.~Majumdar and R.~Tedrake, ``Funnel libraries for real-time robust feedback motion planning,'' \emph{The International Journal of Robotics Research}, vol.~36, no.~8, pp. 947--982, 2017.

\bibitem{Althoff2013}
M.~Althoff, ``Reachability analysis of nonlinear systems using conservative polynomialization and non-convex sets,'' in \emph{Proceedings of the 16th International Conference on Hybrid Systems: Computation and Control}, ser. HSCC '13.\hskip 1em plus 0.5em minus 0.4em\relax New York, NY, USA: Association for Computing Machinery, 2013, p. 173–182.

\bibitem{Murray1984}
D.~M. Murray and S.~J. Yakowitz, ``Differential dynamic programming and newton's method for discrete optimal control problems,'' \emph{Journal of Optimization Theory and Applications}, vol.~43, no.~3, pp. 395--414, 1984.

\bibitem{Tassa2014}
Y.~Tassa, N.~Mansard, and E.~Todorov, ``Control-limited differential dynamic programming,'' in \emph{2014 IEEE International Conference on Robotics and Automation (ICRA)}, 2014, pp. 1168--1175.

\bibitem{Mayne2000}
D.~Mayne, J.~Rawlings, C.~Rao, and P.~Scokaert, ``Constrained model predictive control: Stability and optimality,'' \emph{Automatica}, vol.~36, no.~6, pp. 789--814, 2000.

\bibitem{Mayne2005}
D.~Mayne, M.~Seron, and S.~Raković, ``Robust model predictive control of constrained linear systems with bounded disturbances,'' \emph{Automatica}, vol.~41, no.~2, pp. 219--224, 2005.

\bibitem{Ames2019}
A.~D. Ames, S.~Coogan, M.~Egerstedt, G.~Notomista, K.~Sreenath, and P.~Tabuada, ``Control barrier functions: Theory and applications,'' in \emph{2019 18th European Control Conference (ECC)}, 2019, pp. 3420--3431.

\bibitem{Jaitly2024}
A.~Jaitly and S.~Farzan, ``{PAAMP}: Polytopic action-set and motion planning for long horizon dynamic motion planning via mixed integer linear programming,'' in \emph{2024 IEEE/RSJ International Conference on Intelligent Robots and Systems (IROS)}, 2024, pp. 7617--7624.

\end{thebibliography}

\end{document}